\UseRawInputEncoding
\documentclass[prb,aps,floatfix,twocolumn,superscriptaddress]{revtex4-2}
\usepackage{amsmath,graphics,epsfig,color}
\usepackage{blkarray}
\usepackage{xparse,amssymb,amsfonts,tensor,bm}
\usepackage{xr-hyper}
\usepackage[bookmarks=false]{hyperref} 
\graphicspath{ {./figs/} }

\let\oldhat\hat

\renewcommand{\vec}[1]{\bm{#1}}
\renewcommand{\hat}[1]{\oldhat{\bm{#1}}}
\def\qqc{q}
\DeclareDocumentCommand{\qq}{ O{} O{} }{\vec \qqc_{#1}^{#2}}

\DeclareDocumentCommand{\lmat}{ O{} }{\hat a_{#1}} 

\DeclareDocumentCommand{\ctrans}{ O{} }{\vec t_{#1}} 
\DeclareDocumentCommand{\qdisp}{ O{} O{} }{ \dispsym_{#1}^{#2}}
\DeclareDocumentCommand{\qdispvec}{ O{\cqvec} O{a} O{\alpha} }{ \qdisp[#1][(#2,#3)]}
\newcommand{\dispsym}{u}

\DeclareDocumentCommand{\tset}{ O{} }{\bm{\mathcal T}_{#1}}
\DeclareDocumentCommand{\ssupa}{O{}}{\hat s_{#1}} 
\DeclareDocumentCommand{\phin}{ O{\tset}}{\Phi_{#1}}  
\DeclareDocumentCommand{\supa}{ O{}}{\hat S_{#1}}
\DeclareDocumentCommand{\bqset}{ O{} O{}}{\tilde Q_{#1}^{#2}} 
\DeclareDocumentCommand{\dqn}{ O{\qset} }{D_{#1}}  
\DeclareDocumentCommand{\qset}{ O{} }{\vec Q_{#1}} 

\NewDocumentCommand{\ceil}{s O{} m}{%
\IfBooleanTF{#1} 
{\left\lceil#3\right\rceil} 
{#2\lceil#3#2\rceil} 
}
\begin{document}

\title {Benchmarking phonon anharmonicity in machine learning interatomic potentials}
\author{Sasaank Bandi}
\address{Department of Applied Physics and Applied Mathematics, Columbia University, New York, NY 10027, USA}
\author{Chao Jiang}
\address{Computational Mechanics and Materials Department, Idaho National Laboratory, Idaho Falls, ID 83415, USA.}
\author{Chris A. Marianetti}
\address{Department of Applied Physics and Applied Mathematics, Columbia University, New York, NY 10027, USA}
\begin{abstract}

Machine learning approaches have recently emerged as powerful tools to probe
structure-property relationships in crystals and molecules. Specifically,
Machine learning interatomic potentials (MLIP) can accurately reproduce
first-principles data at a cost similar to that of conventional interatomic potential
approaches. While MLIP have been extensively tested across various classes of
materials and molecules, a clear characterization of the anharmonic terms
encoded in the MLIP is lacking. Here, we benchmark popular MLIP using the
anharmonic vibrational Hamiltonian of ThO$_2$ in the fluorite crystal
structure. This anharmonic Hamiltonian was constructed from density functional
theory (DFT)  using our highly accurate and efficient irreducible derivative
methods, and then used to generate molecular dynamics trajectories. This data
set was used to train three classes of MLIP: Gaussian Approximation Potentials,
Artificial Neural Networks (ANN), and Graph Neural Networks (GNN). The results
were assessed by directly comparing phonons and their interactions, as well as
phonon linewidths, phonon lineshifts, and thermal conductivity. The models were
also trained on a DFT molecular dynamics dataset, demonstrating good agreement
up to fifth-order for the ANN and GNN. Our analysis demonstrates that MLIP have
great potential for accurately characterizing anharmonicity in materials
systems at a fraction of the cost of conventional first principles-based
approaches.
\end{abstract}

\maketitle

\section{Introduction}
Phonon anharmonicity plays an essential role in the thermodynamics of materials
systems. Thus, constructing accurate vibrational Hamiltonians is critical for
materials property prediction at finite temperatures. 
In particular, faithfully resolving phonon interactions up to fourth-order is essential
for capturing leading order behavior of the phonon linewidths, phonon lineshifts, and
thermal conductivity. In order to systematically assess the phonons and their interactions,
it is critical to use the minimum set of derivatives allowed by group theory,
which may be computed using irreducible derivative approaches \cite{Fu2019,Mathis2022,Bandi2023,Xiao2023}. 
Though these irreducible derivative methods are significantly more efficient
than competing finite difference approaches, higher-order derivatives in
materials with large supercells may still be prohibitively expensive. 
Recently, machine learning interatomic
potentials (MLIP) have emerged as powerful tools to represent the Born-Oppenheimer potential
of materials. MLIP methods can achieve near quantum chemical
accuracy relative to the \textit{ab initio} method they were trained on while avoiding the polynomial scaling \cite{Batzner2022}. Here, we assess the fidelity of anharmonic derivatives 
computed from MLIP.

MLIP aim to represent the total
energy of a given atomic configuration in terms of individual contributions
from each atom. Since the complete space of atomic
displacements is high dimensional for very large molecules and
crystals, several descriptors have been developed to encode local atomic
environments \cite{Langer2022} such as 
coulomb matrices \cite{Rupp2012}, atom
centered symmetry functions \cite{Behler2011}, smooth overlap of atomic
positions \cite{Bartok2013}, and
histograms of distances, angles, and
dihedrals (HDAD) \cite{Faber2017}. 
While understanding how well certain descriptor
sets can represent anharmonic energy surfaces is important, it will not be
directly discussed in this work and will be the subject of further study. 
Our present goal is to assess the most prevalent MLIP in the literature,
using the commonly associated descriptors.
Therefore, our benchmarks will be limited to three popular classes of potentials with
publicly available software. 

The first MLIP considered in this work is the Gaussian
approximation potential (GAP) \cite{Bartok2010}. Here, each atomic energy is expanded in terms of
basis functions of the descriptors, and the distribution of basis weights is
assumed to be normally distributed about zero. Therefore, the energy of a given
configuration is also normally distributed, where the mean provides a prediction
from the model, and the variance can be a measure of
uncertainity. The advantages of this type of potential are the
characterization of model uncertainty and that the analytical form of the prediction
is a sum of basis functions. In practice, the size of basis functions may be
prohibitively large, and the ``kernel trick'' is employed to reduce memory and
cost requirements. It should be noted that the energy prediction provided by GAP
is equivalent to the prediction of the kernel ridge regression on the same
basis set. Recently, GAP has been used to predict second and third-order
derivatives in two dimensional materials where good agreement with DFT is
achieved with integrated properties such as phonon lifetimes and
thermal conductivity
\cite{Kocabas2023,Ouyang2020,Zhang2020,Wei2022,Arabha2021}. Online trained GAP models have
also been implemented in the Vienna \textit{ab initio} software package
(VASP) \cite{Kresse1996,Kresse1993} and have also been used along with
approaches such as the stochastic self-consistant harmonic approximation (SSCHA) \cite{Monacelli2021} to predict temperature dependent phonon properties
using monte carlo sampling \cite{Ranalli2023,Verdi2023}. 

While not directly formulated as a MLIP, several potential energy surface
fitting approaches, such as compressive sensing lattice dynamics (CSLD)
\cite{Zhou2014,Zhou2019}, have a similar mathematical foundation as GAP. In both
models, a Bayesian polynomial regression framework is used to fit model parameters to sampled DFT data. 
While GAP generally fits descriptors of the local atomic
environment, CSLD uses the complete Cartesian coordinate basis which ensures that
fitting coefficients are more analogous to force constants. Due to the combinatorial explosion of
high-order derivatives when using the Cartesian basis, least squares solutions
tend to overfit parameters. CSLD overcomes this impediment by fitting force constants
using the least absolute shrinkage and selection operator (LASSO), which has
the same objective function as ordinary least squares but with an additional penalty on the $L_1$
norm of the model parameters \cite{Tibshirani1996}. The sparse
solutions provided by LASSO have yielded accurate anharmonic properties in various systems
\cite{Zhou2014,Tadano2015,Zhou2019}. 

The second class of machine learning approaches we consider is the high dimensional
artificial neural network first described by Behler and
Parrinello (BPNN) \cite{Behler2007}. In this model, each atomic species is
represented by a convolutional neural network, and the total energy is computed
as a sum of the output of each neural network. Historically, atom-centered
symmetry functions \cite{Behler2011} have been used as descriptors to encode
rotational invariance and have been systematically improved over time.
Recently, a set of descriptors constructed as irreducible representations of the
Euclidean group has been shown to encode rotational equivariance, providing 
better predictions of tensor properties \cite{Geiger2022}, such
as the forces. BPNNs have been used to predict phonon dispersions and thermal
conductivity in semiconductors \cite{Li2020,Li2020a,Mangold2020,Minamitani2019},
new thermoelectric candidate materials \cite{Choi2022}, and
superlattices \cite{Qu2023}.

Finally, the most recent developments in machine learning interatomic
potentials have focused on graph neural
networks \cite{Batzner2022,Takamoto2022,Batatia2022,Gasteiger2020,Chen2022,Deng2023}. In
this architecture, graphs are constructed for all neighboring atoms where nodes
encode atomic information, and bonding information is incorporated through edge
connections. Of the three models described in this work, graph neural networks
have been shown to achieve the highest accuracy in reproducing forces and
energies while requiring fewer training samples than the other two approaches.
Recently, several developments have been aimed at developing a universal
interatomic potential trained on large DFT databases, such as the materials
project \cite{Jain2013}, and phonon dispersions computed using these universal models have shown
good agreement with DFT \cite{Takamoto2022,Chen2022,Deng2023}. While graph neural networks have been used to predict
thermal conductivity directly by training on
experimental and first principles data from materials property databases \cite{Ojih2023,Sun2021}, we
are not aware of any other work using graph neural network interatomic
potentials to compute anharmonic observables via phonon interactions.

In this work, we benchmark how well MLIP
reconstructs anharmonic potentials. We trained the GAP,
BPNN, and GNN using two datasets: one containing configurations evaluated by an anharmonic Taylor series, which does not contain noise, 
and one generated using DFT calculations.
In our analysis, we quantify errors at the level of the irreducible derivatives
of the potential energy surface, making our benchmark 
more robust than benchmarks using integrated observables. In all three models,
third-order derivatives are in good agreement with the reference. Up to fifth-order derivatives are mostly reproduced by the ANN and GNN, with the GNN
demonstrating particularly promising accuracy. We also provide comparisons of
observables such as phonon dispersions, phonon linewidths, phonon lineshifts, and
thermal conductivity to
quantify the effects of the errors in the irreducible derivatives.

\section{Computational Methods and Details}
DFT calculations were performed using the projector augmented wave (PAW) method
implemented in the Vienna \textit{ab initio} software package (VASP) using the local
density approximation \cite{Kresse1996,Kresse1993}. The plane-wave basis cutoff energy was set to 500 eV. A
$\Gamma$ centered 10$\times$10$\times$10 $k$-point mesh was used for primitive cell calculations,
and measurements in other supercells were conducted with corresponding $k-$mesh
denisties. Gaussian smearing with $\sigma=0.1$ was used to avoid numerical
errors during $k$-point summation. DFT energies were converged to within
$10^{-6}$ eV, and the unit cell was relaxed until all forces were within $0.02$
eV/\AA.

Second (phonons), third, fourth, fifth, and sixth-order derivatives of the
Born-Oppenheimer potential of ThO$_2$ were computed from density functional theory
and the MLIP via using irreducible derivative approaches \cite{Fu2019}. The finite difference
calculations were extrapolated to a discretization
size of zero using quadratic errortails with ten $\Delta$s (for more information see \cite{Fu2019,Bandi2023}).
Phonons and third-order derivatives commensurate with the $2\times2\times2$ supercell,
which contains 24 atoms, were computed  using the lone irreducible derivative approach with forces
(LID$_1$) and energy derivatives (LID$_0$) for DFT and
MLIP, respectively. Fourth-order phonon interactions commensurate with the conventional cubic supercell, which contains 12 atoms, were computed
using the bundled irreducible derivative (BID) approach for DFT due to the large
computational cost. While BID generally evaluates derivatives using the
smallest set of measurements allowed by group theory, in this case the number
of measurements was tripled to ensure robust derivatives. The fourth-order derivatives were evaluated with LID$_1$ for all three
machine learning models. Fifth-order derivatives associated with the following $Q$ points
were selectively computed using LID$_1$ for both DFT and MLIP:
$Q =
\left(\left(\frac{1}{2},0,0\right),\left(\frac{1}{2},0,0\right),\left(0,0,0\right),\left(0,0,0\right),\left(0,0,0\right)\right)$
and $Q =
\left(\left(0,0,0\right),\left(0,0,0\right),\left(0,0,0\right),\left(0,0,0\right),\left(0,0,0\right)\right)$
Finally, only sixth-order interactions
commensurate with the primitive cell were measured with LID$_1$. 

Two datasets were prepared to train the machine learning models. 
The first dataset derives from a vibrational Hamiltonian containing
second, third, and fourth-order irreducible derivatives computed from DFT.
Datapoints are generated by performing molecular dynamics on the irreducible
derivative Hamiltonian using the microcanonical ensemble, which we refer
to as irreducible derivative molecular dynamics (IDMD) \cite{Xiao2023}.
As a
result, all energies and forces in the dataset are in perfect
agreement with the irreducible derivatives, yielding a 
noiseless dataset, which we refer to as the \textit{IDMD} dataset.
The IDMD was conducted in the $4\times4\times4$
supercell, containing 192 atoms, for $4000$ steps with a timestep of $4$fs. MD
trajectory velocities were initialized by sampling from the Maxwell-Boltzmann
distribution at 2000K and a temperature range of 1522K-2238K was observed
throughout the simulation. The second dataset, referred to as
\textit{DFTMD}, was generated using \textit{ab initio} molecular dynamics in the conventional cubic
cell tripled in all three dimensions ($3S_C$), which contains 324 atoms. MD was conducted using the canonical
ensemble at 2000K and 3000K, and 1346 snapshots were taken along the two trajectories.
While this dataset contains noise inherent to the numerics of the DFT calculation, it probes all
derivatives commensurate with the $3S_C$ supercell up to infinite-order. Each
dataset was constructed to provide a unique test of the MLIP. Since
\textit{IDMD} is a noiseless dataset, this benchmark 
purely probes each MLIP ability to learn anharmonic interactions
while avoiding any influence of how well the model accuracy scales with noise.
On the other hand, \textit{DFTMD} is a dataset that has been constructed
in accordance with the current state-of-the-art and represents a typical dataset
used for MLIP training. Since \textit{DFTMD} dataset contains infinite-order interactions, it provides the
opportunity to test the limits of each MLIP and determine the order at which
the MLIP fails to replicate DFT results.

All three MLIP were trained on \textit{IDMD}, and only
the BPNN and GNN were trained on \textit{DFTMD}.
The GAP model was generated using the quantum mechanics and interatomic
potentials (QUIP) code \cite{Csanyi2007,Kermode2020,Bartok2010}. Descriptors
were constructed using two-body and three-body terms with a cutoff of 6 \AA,
along with the smooth overlap of positions (SOAP) descriptor with a cutoff of 4
\AA. The neural network potential was trained using the n2p2 code
\cite{Singraber2019,Behler2007}.  Atom-centered radial and angular symmetry
functions with a cutoff radius of 10\AA{} and 6\AA{} were used, respectively.
For each atom, a four-layer neural network was constructed which contained two
hidden layers with 21 neurons each. The size of the input layer for thorium
atoms was 41, while the size was 46 for oxygen atoms. Finally, the graph neural
network was trained using nequIP \cite{Batzner2022}. Graph edges were
constructed with a cutoff of 6 \AA{} with four interaction blocks. The E(3)
equivariant spherical harmonics were used with radial basis functions to form
descriptor sets \cite{Geiger2022}. More detailed model information for all
three machine learning interatomic potentials is provided in supplementary
information.

\section{results}
\begin{figure}[h]
  \begin{center}
    \includegraphics[width=0.95\linewidth]{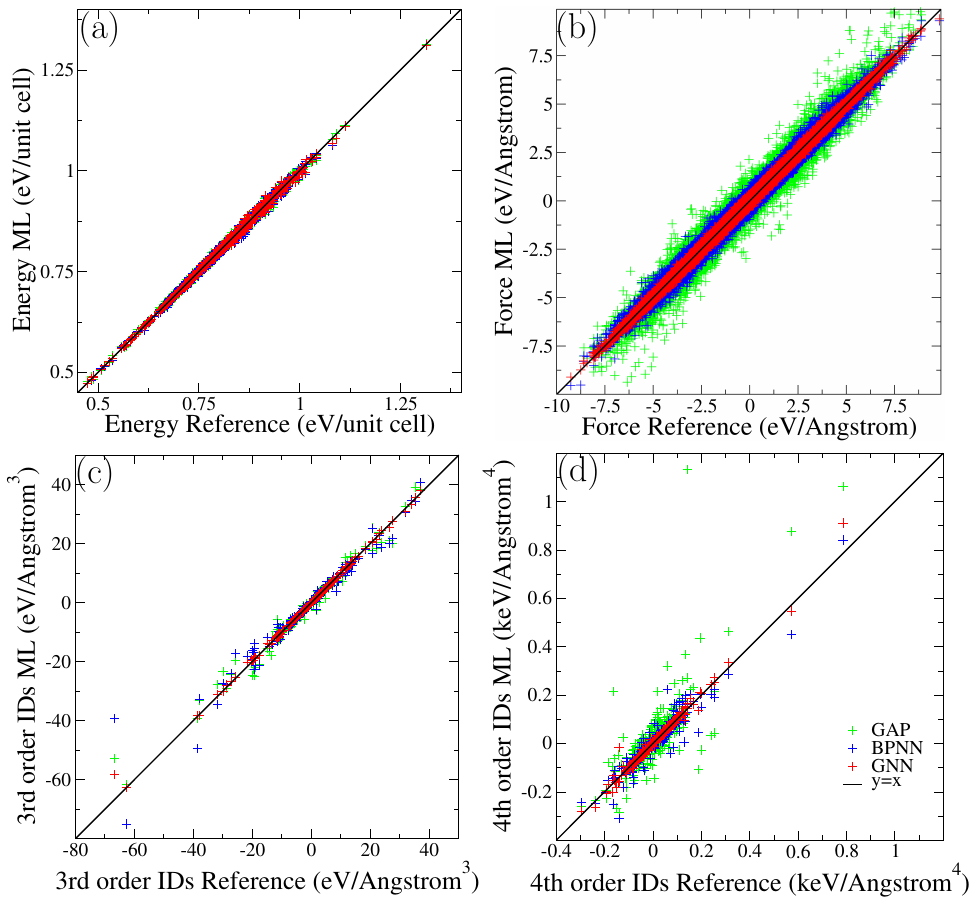}
  \end{center}
  \caption{Results of the MLIP trained on \textit{IDMD} compared to
    the reference, where green,
  blue, and red dots indicate GAP, BPNN, and GNN results, respectively. Panels $a$
and $b$ show the energies and forces for a distinct testing set, respectively. Panels $c$
and $d$ compare the  third and fourth-order irreducible derivatives computed from MLIP to the
reference irreducible derivatives, respectively.}
\label{fig:IDMD}
\end{figure}

\begin{figure}[h]
  \begin{center}
    \includegraphics[width=0.95\linewidth]{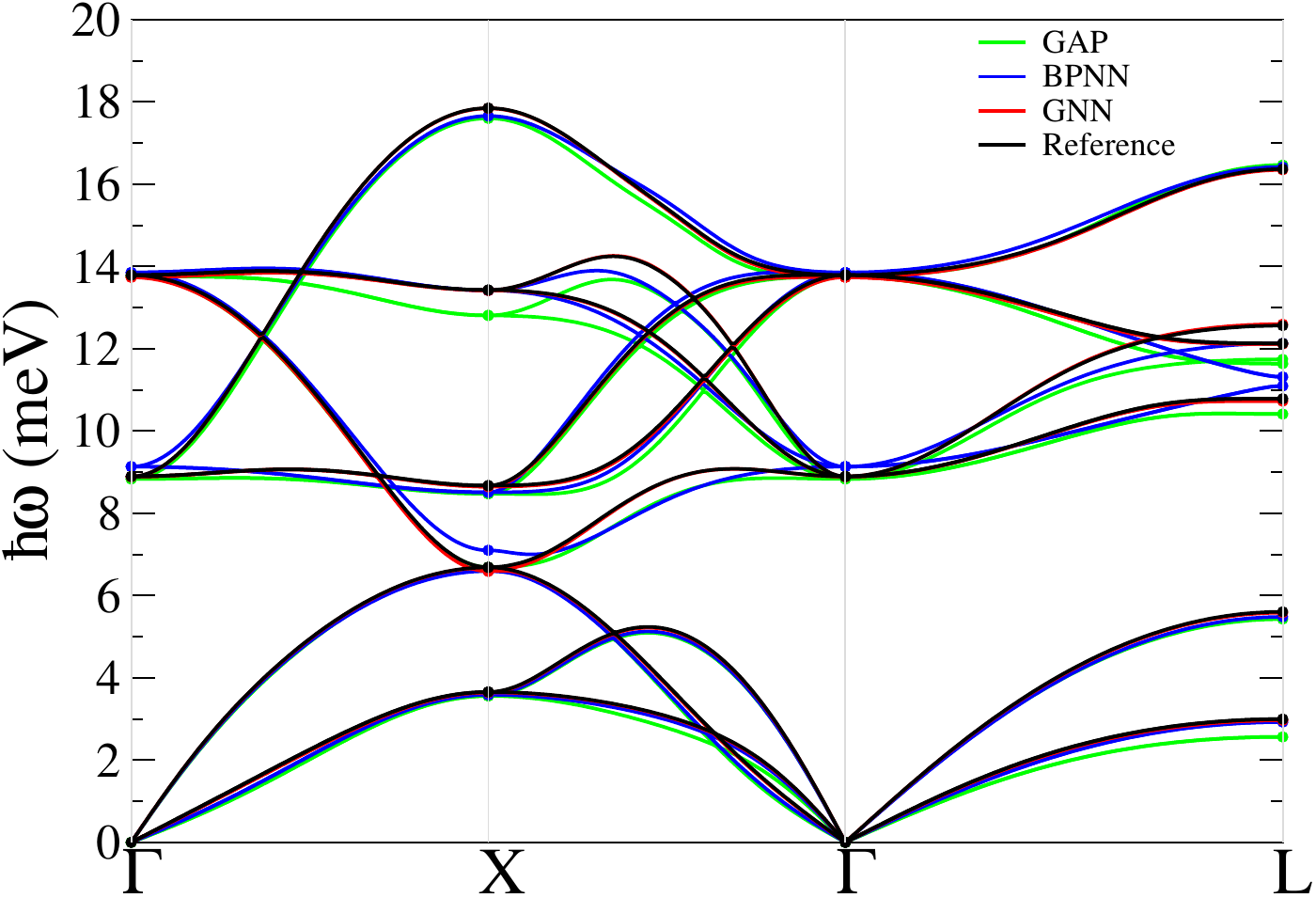}
  \end{center}
  \caption{Phonon dispersion computed in the $2\hat I$ FTG, where green, blue, red, and
  black points represent explicit measurements of the irreducible derivatives using
GAP, BPNN, GNN, and the reference, respectively. Solid lines are a Fourier interpolation.}
\label{fig:phonon}
\end{figure}

\begin{figure}[h]
  \begin{center}
    \includegraphics[width=0.95\linewidth]{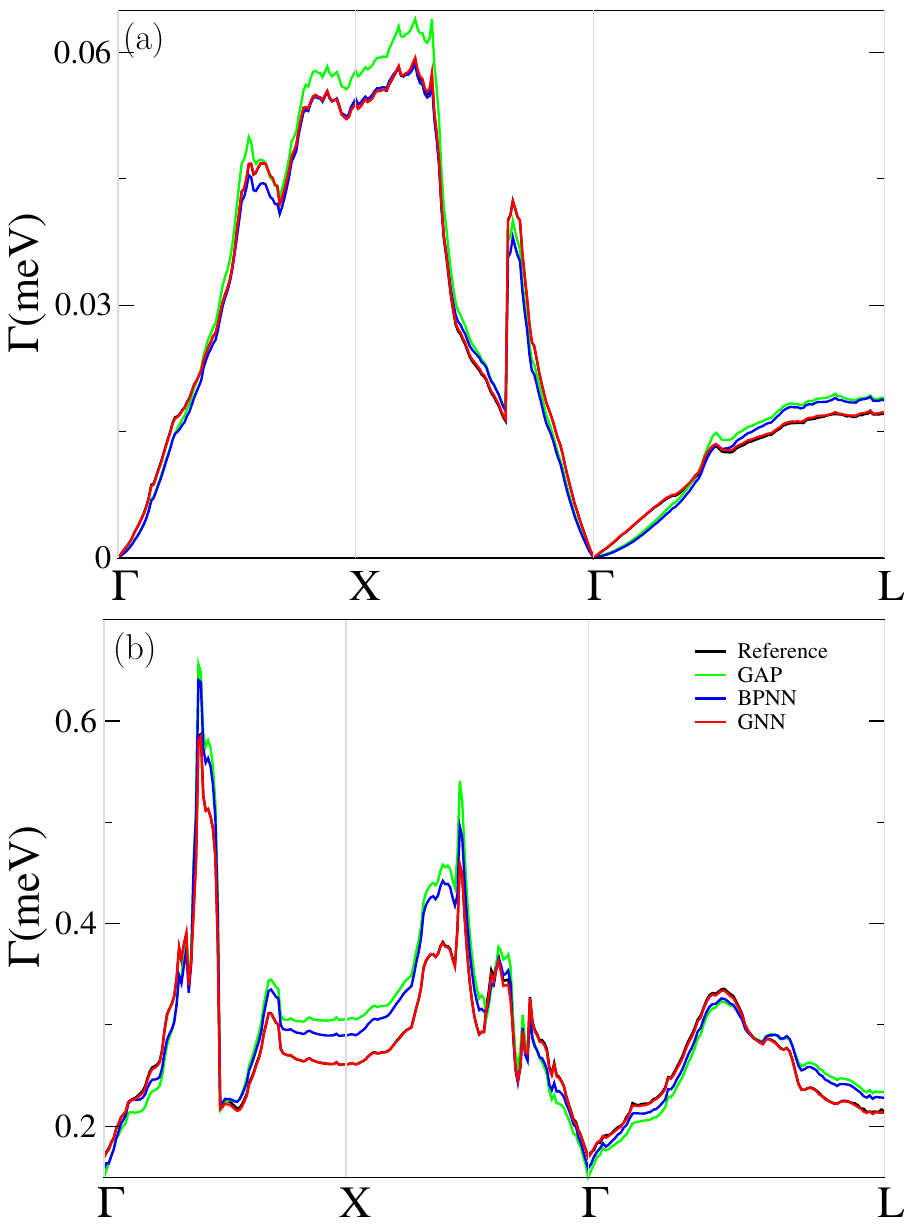}
  \end{center}
  \caption{Phonon linewidth contribution from the bubble diagram, which uses the third-order irreducible derivatives. Panels $a$ and $b$ show the linewidth of an acoustic
  and optical branch along a high symmetry path, respectively.  
  The GAP, BPNN, GNN, and the reference are shown in green, blue, red, and
black, respectively.}
\label{fig:linewidth}
\end{figure}
\begin{figure}[h]
  \begin{center}
    \includegraphics[width=0.95\linewidth]{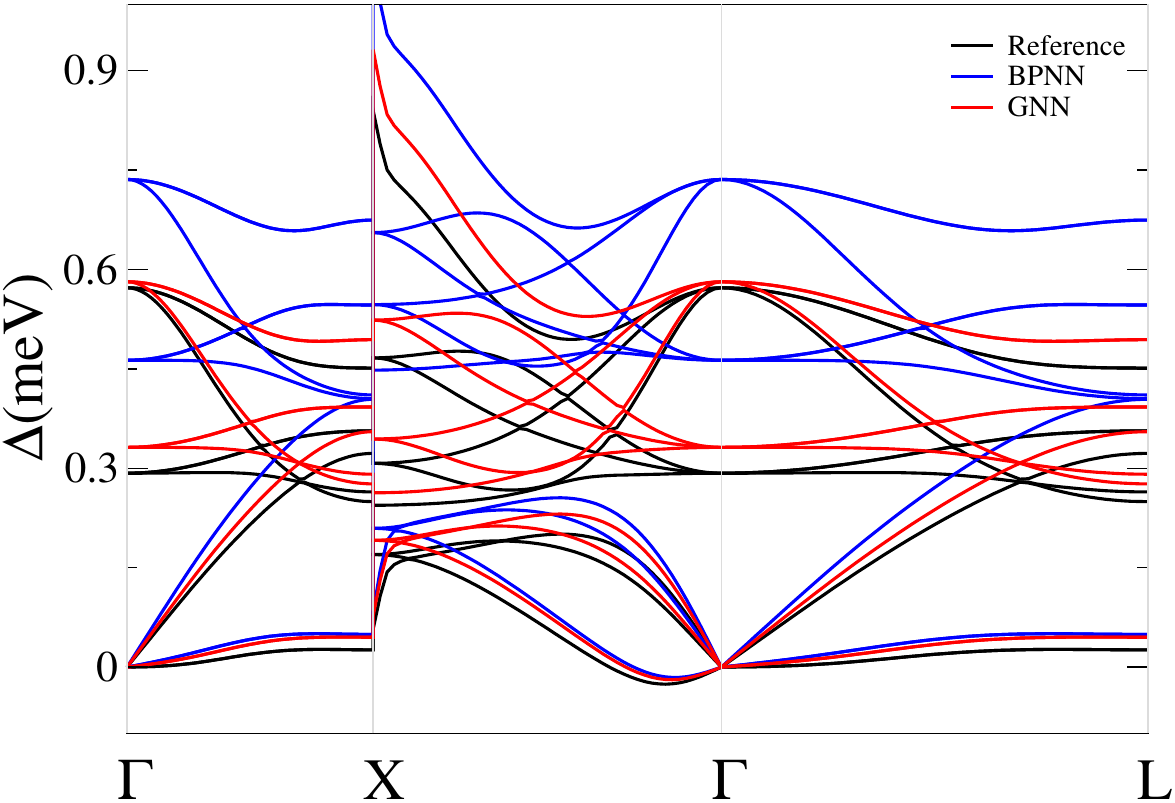}
  \end{center}
  \caption{Phonon lineshift contribution from the loop diagram, which uses the fourth-order irreducible derivatives. The BPNN, GNN, and the reference results are shown in blue,
red, and black, respectively.}
\label{fig:deltas}
\end{figure}

\begin{figure}[h]
  \begin{center}
    \includegraphics[width=0.95\linewidth]{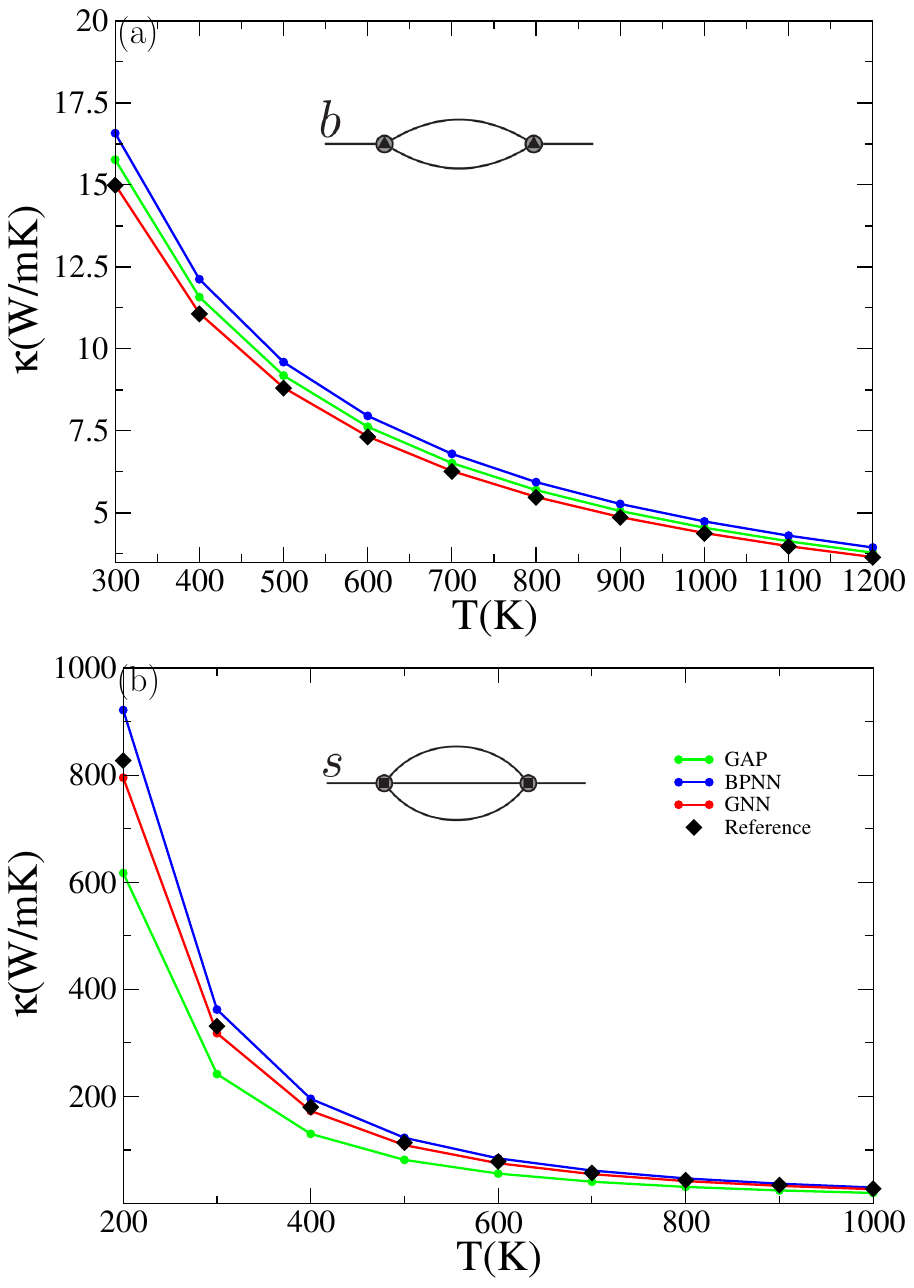}
  \end{center}
  \caption{Thermal conductivity evaluated using lifetimes from
  the bubble diagram in panel $a$ and lifetimes exclusively from the  
sunset diagram in panel $b$. GAP, BPNN, and GNN results are shown with green,
blue, and red solid dots and the reference results are shown with black diamonds. Solid
lines are drawn to direct the eyes.}
\label{fig:tc}
\end{figure}

\begin{figure}[h]
  \begin{center}
    \includegraphics[width=0.95\linewidth]{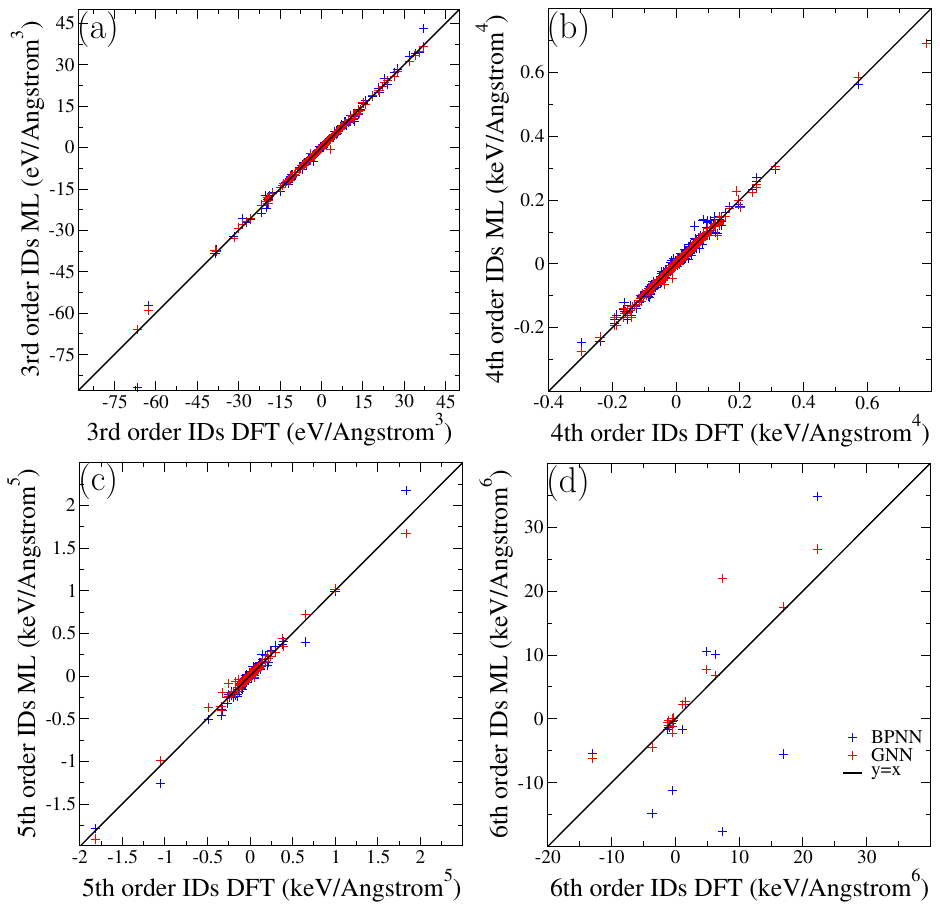}
  \end{center}
  \caption{Anharmonic derivatives of the MLIP trained on \textit{DFTMD}
  compared to the DFT reference where BPNN and GNN results are shown in blue and
red, respectively.}\label{fig:DFT}
\end{figure}
\begin{table*}[t]
\begin{tabular}{ccclcccc}
\hline
\multicolumn{1}{l}{}    & \multicolumn{2}{c}{\textit{IDMD}} &  & \multicolumn{4}{c}{\textit{DFTMD}}         \\
                        & 3rd Derivative                & 4th Derivative                &  & 3rd Derivative & 4th Derivative & 5th Derivative & 6th Derivative \\ \cline{2-3} \cline{5-8} 
GAP                     & 0.2207                        & 1.652                         &  & -              & -              & -              & -              \\
BPNN                    & 0.3100                        & 0.480                         &  & 0.1853         & 0.1890         & 0.4707         & 0.7965         \\
\multicolumn{1}{l}{GNN} & 0.0747                        & 0.212                         &  & 0.0587         & 0.1346         & 0.2991         & 0.1359         \\ \hline
\end{tabular}
\caption{Table of root mean squared error divided by the average magnitude of the irreducible derivatives for each model trained on the \textit{IDMD} and \textit{DFTMD} datasets.}
\label{tab:total}
\end{table*}

We begin by analyzing the results of the machine learning models trained on
\textit{IDMD}. In Fig. \ref{fig:IDMD}, we show
the results of all three models where GAP, BPNN, and GNN results are shown in
green, blue, and red, respectively. In figure \ref{fig:IDMD} ($a$), we
compare energies computed on a distinct testing set generated using IDMD, showing excellent
agreement. Figure \ref{fig:IDMD} ($b$) compares the forces from all three models
on the testing set; the forces from GAP are in reasonable agreement with the reference,
while the BPNN and GNN are in much better agreement. In panels \ref{fig:IDMD}($c$) and \ref{fig:IDMD}($d$)
and table \ref{tab:total}, the third and fourth-order irreducible derivatives
are compared to the reference. In all three models, the third-order
derivatives are in good agreement with the reference, with the GNN yielding the
best fidelity of the three. At fourth-order, the BPNN and GNN are in reasonable
and good agreement with the reference, respectively, while the GAP model fails to provide
accurate irreducible derivatives. 

To demonstrate the practical impact of the errors in the computed irreducible
derivatives, we compare several observables computed using irreducible derivatives
obtained from each model. We begin by presenting the phonon dispersion (see
Fig. \ref{fig:phonon}). Solid dots represent grid points where irreducible
derivatives are computed, and the solid lines are a Fourier interpolation.
We demonstrate that for most phonon branches, all three models are in good
agreement with the reference, though the BPNN and the GAP both have non-trivial
discrepancies at the $X$ and $L$ points. We proceed by presenting the phonon
linewidths from the bubble diagram \cite{Xiao2023}, which uses the third-order irreducible derivatives, in a typical acoustic and optical branch (see Fig.
\ref{fig:linewidth}). All other branches are included in supplementary material. In the acoustic branch (Fig \ref{fig:linewidth}($a$)), all
three models are in relatively good agreement with the reference, though GAP overestimates
the linewidths near the $X$ point by about $16\%$. The optical branch linewidths
(Fig. \ref{fig:linewidth}($b$)) are more sensitive to errors in the irreducible
derivatives, and thus, both GAP and BPNN yield discrepancies up to $30\%$ between
$\gamma$ and $X$. We note that the GNN yields near perfect
agreement with the reference for both branches. Next, we present the phonon lineshifts computed by
evaluating the loop diagram \cite{Xiao2023} using the fourth-order irreducible derivatives (see
Fig. \ref{fig:deltas}). Only the BPNN and GNN lineshifts have been plotted
due to the massive discrepancies in the GAP results (see supplementary material). While both
models are in good agreement with the reference for the acoustic branches, the BPNN
significantly overestimates the shifts of the optical branches, while the GNN is
still within $18\%$ of the reference. Finally, we compare the thermal
conductivity computed using the phonon lifetimes in the Boltzmann
transport equation within the relaxation time approximation. In panel
\ref{fig:tc}($a$), the phonon lifetimes have been evaluated using the bubble diagram,
which only uses the third-order derivatives, while in panel \ref{fig:tc}{$b$}
phonon lifetimes have been evaluated using only the sunset diagram \cite{Xiao2023}, which 
uses the fourth-order irreducible derivatives. In both cases, it is evident that the
thermal conductivity is less sensitive to errors in the irreducible derivatives,
and all three models are in good agreement with the reference.

We proceed by discussing the results of the BPNN and GNN models trained on
\textit{DFTMD}. Since we have already demonstrated that the GAP model
fails to capture interactions beyond third-order, it was not used for this
benchmark aimed at determining the highest order of derivatives that can be
faithfully reproduced using MLIP. We present a comparison of the third, fourth,
fifth, and sixth-order irreducible derivatives computed from the MLIP and DFT
(see Fig. \ref{fig:DFT} and Table \ref{tab:total}). For third
\ref{fig:DFT}($a$), fourth \ref{fig:DFT}($b$), and fifth-order
\ref{fig:DFT}($c$) derivatives, both models are in excellent agreement with DFT,
with the GNN outperforming the BPNN by a factor of $1.5$ to $3$ depending
on the derivative order. At sixth-order \ref{fig:DFT}($d$), both models fail to reproduce the
irreducible derivatives computed from DFT, though the graph neural network
generates reasonable results, especially on low-magnitude derivatives.

\section{Conclusions}
In summary, we have benchmarked three popular machine learning interatomic
potentials on the anharmonic irreducible derivatives of ThO$_2$. We have developed two training datasets:
one that is computed from the anharmonic vibrational Hamiltonian  of ThO$_2$, containing up to quartic terms, and one that is representative of training conducted
via \textit{ab initio} methods in the literature. The Behler Parrinello
artificial neural network and the graph neural network yield robust
irreducible derivatives up to fifth-order, while the Gaussian
approximation potential is only able to accurately capture anharmonic
interactions up to third-order. Our work demonstrates the promising potential of
machine learning methods in characterizing anharmonicity in materials systems.
While we obtained accurate results using MLIP in the present study, it would be interesting
to verify that our findings hold in systems with sensitivities, such as soft phonon modes.
Future work will include extending this analysis to strongly correlated electronic
materials, where the generation of accurate training data comes at a
significant premium. Additionally, to study materials with strong anharmonicity, further work will be conducted to develop
training algorithms to more accurately capture sixth-order and beyond
interactions.
\section{Acknowledgements}
The development of MLIP for ThO$_2$ was supported by the INL Laboratory Directed
Research \& Development (LDRD) Program under DOE Idaho Operations Office
Contract DE-AC07-05ID14517.  The construction of anharmonic Hamiltonians was
supported by the grant DE-SC0016507 funded by the U.S. Department of Energy,
Office of Science.  This research used resources of the National Energy
Research Scientific Computing Center, a DOE Office of Science User Facility
supported by the Office of Science of the U.S. Department of Energy under
Contract No. DE-AC02-05CH11231.

\bibliography{main}
\end{document}